\documentclass[allclo,superscriptaddress,eqsecnum,amsfonts,showpacs]{revtex4}
\usepackage{epsfig}

\newcommand{\be}{\begin{equation}}
\newcommand{\ee}{\end{equation}}
\newcommand{\bea}{\begin{eqnarray}}
\newcommand{\eea}{\end{eqnarray}}

\newcommand{\ep}{i\varepsilon}
\newcommand{\nn}{\nonumber}

\newcommand{\om}{\omega}

\begin{document}

\preprint{ \parbox{1.5in}{\leftline{hep-th/??????}}}

\title{Confined gluon from  Minkowski space continuation of PT-BFM  SDE solution}

\author{Vladimir ~\v{S}auli}
\affiliation{DTP INP Rez, CAS. }

\begin{abstract}

Recent lattice studies exhibit infrared finite effective QCD charges. Corresponding gluon propagator in Landau gauge is finite and nonzero, suggesting a mechanism of dynamical gluon mass generation is in the operation. In this paper, the analytical continuation of the Euclidean (spacelike) Pinch Technique-Background Field Method (PT-BFM) solution of Schwinger-Dyson equation for gluon propagator to the timelike region of $q^2$ is found.  We found the continuation numerically showing good agreement   with a generalized Lehman representation for small Schwinger coupling. The associate non-positive spectral function has an unexpected behavior. Albeit infrared Euclidean space solution naively suggests like single scale "massive"  propagator, the obtained spectrum of gluon propagator does not correspond  to the delta function at single scale $q=m$ , instead  more possible singularities are generated. The pattern depends on the details of assumed Schwinger mechanism: for stronger coupling there are few maxima and minima which appear at the scale $\Lambda$, while for perturbatively small Schwinger coupling the spectral function shows up two narrow peaks: particle and ghost  excitation, which have mutually opposite signs.  
\end{abstract}

\maketitle

%-------------------------------------------------------------------------------------------------------

\section{Introduction}

Infrared behavior of Green's functions (GFs) of pure Yang-Mills theories has been intensively studied in the last decade.
Even though GFs are gauge fixing dependent objects and thus they do not represent physical observables, it is believed they include reliable   nonperturbative information about confinement (and chiral symmetry breaking when quarks are included). 
Various solutions for GFs have been obtained by solving Schwinger-Dyson equations (SDEs) and by  lattice calculations.
Recently, study of SDEs offer two scenarios which are distinguished by the infrared behavior  of GFs.  The so called  scaling solution exhibits power law momentum behavior with an infrared exponents  \cite{LESM2002,ZW2002,FIPA2007} which lead to infrared vanishing Landau gauge gluon propagator $\Delta(0)=0$ and correspondingly infrared enhanced ghost propagator. For a SDEs treatment in QCD  let us refer \cite{ROBSCM,ALKSME2001}.

 On the other side, recent lattice calculations \cite{BOU2007,BLYMPQ2008,lat1,lat2,lat3,lat4,CUME2008,BIMS2009,IRITANI,DUGRSO2008,DUOLVA2010,BGLYMRQ2010,TW2010} in conventional Landau gauge support the so called {\it decoupling} solution.
 In this scenario the gluon propagator $\Delta(0)\not= 0$, but stays finite. Also, it is notable, the ghost propagator is not enhanced in the infrared and thus remains semi-perturbative. Such decoupling solutions, called sometimes  massive solutions were already proposed in  \cite{COR1982,COR1986,COR1989} and being based on attractive theoretical frameworks of  Pinch Technique and Background Field Method (PT-BFM) they  attract many new attentions these days  \cite{JHEP,PENIN,BIPA2002,BIPA2004,BIPA2008,cornwall2009,SA2010,AGBIPA2008,AGBIPA2011}. Comparing to standard gauge fixed set of GFs it has been proved that using Pinch Technique- one can rearrange usual gauge fixed GFs (in any gauge) in a way that they do not depend on the gauge fixing parameters and PT GFs satisfy original WTI an not more complicates Slavnov-Taylor identities, for the topical review see \cite{rewiev2009}.
While scaling solution would be compatible with Gribov-Zwanziger confinement scenario, it is believed that the decoupling solutions  is related to a dynamical generated gluon mass $m^2 \simeq \Delta^{-1}(0)$, however  confinement mechanism of associated massive gluons remains to be determined.

The truncation of such reorganized SDEs  with WTI satisfying  internal vertices is  called  "PT-BFM inspired" or simply PT-BFM gluon SDE. As a starting point we use the  model originally solved in Euclidean space and for certain set of free  parameters presented in the paper \cite{JHEP}.  The dynamical mass generation  
is a BFM generalization of the well known Schwinger mechanism, hereby singular vertex appears  in quantum loops only and not due to the  Goldstone modes, which do no take place here as the gauge  symmetry is exactly  preserved.
 However as we do not solve more complicated SDE for the gluon vertices,  we freely accommodate modeled transversal part in its original form \cite{JHEP}. Having solved the problem in the Euclidean space we formally perform analytical continuation of the equation to the timelike region, i.e. to the Minkowski space, where it should provide analytical continuation as a  solution.

  In BFM scheme, the expected  Schwinger mechanism gives rise to the infrared gluon mass which naively could not be   so far from the value of $\Delta^{-1/2}(0)$, while the true physical particle mass is determined as the pole mass of  the full propagator. For a stable unconfined particle  it can be extracted as a real solution  of equation: $\Delta^{-1}(q^2=m_p(q^2))=0$. Gluon is experimentally unobservable as a particle and thus very likely free plane wave  solution does not exists. If there is no real pole in the gluon propagator then a propagation of on-shell gluon is indeed impossible. What is the singularity structure of the propagator of expectingly confined excitations  has not been known. Instead of this, some attempts to determine (not completely understood) gluon mass scale were performed on various theoretical and phenomenological basis   \cite{cornwall2009,SA2010,NATALE, NATALE2}
giving us rough and simplified estimate $m_g=\Lambda_{QCD}$.

 In this paper we study Minkowski space continuation of already considered PT-BFM solution \cite{JHEP} ,which although well defined in the entire Minkowski space, was numerically performed in Euclidean space only. 
In reality, we did not get just one  solution of continued PT-BFM gluon SDE but many, among them we have to choose the correct solution.
For this purpose it is fully legitimate to check the  analytical properties which are equivalent to the analyticity  Stieltjes transformable function (generalized Lehman representation).
This is fully consistent treatment as the Lehman representation was explicitly used during the derivation of the original Euclidean space SDE. Due to this reason we will abandon all the solutions which largely  do not fit  dispersion relation dictated by Lehman representation. Optimizing the analyticity as far  possible we got a few scenarios, albeit they are  quite sensitive to the details of the model. Especially, and quite interestingly 
it allows to answer what could be a physical mass of the gluon in the pure Yang-Mills theory. Neither we found  single solution $\Delta^{-1}(q^2=m_p^2)=0$ for some pole mass $m_p$. Actually,  solutions we have found  lie between two extreme cases: first is the solution with  two particle like singularities with opposite signs and the second case, where there is no  solution for the physical mass shell. In later case the associated spectral function is an oscillatory one with a few number of relatively smooth minima and maxima. The later case we can interpret as a spectrum of confined gluon, while the first kind of solutions is very likely an artifact of approximation made.
In both cases the solution suggests the first branch point is located in the origin of complex $q^2$ plane.

Our convention for Minkowski  metric tensor reads: $g_{\mu\nu}=diag(+1,-1,-1,-1)$. Minkowski momenta are not labeled in our notation,  while we always write $E$ when we specify the  Euclidean momenta, i.e. for instance $q^2_E=-q^2$   for some spacelike momentum $q$.  We use simple letter $d$ for the PT-BFM gluon propagator in Minkowski and $d_E$ in Euclidean space, while symbol  $\Delta$ is used for a conventional propagator defined in given gauge. 
In the next Sections II and III we describe necessary  ingredients of the original PT-BFM  SDE and perform formal "analytical" continuation. In Section IV  we continue with RG improved SDEs for which we present numerical solutions. We discuss  what happens to the solutions in various cases, we also discuss necessary amount of numerics, which is the key tool to get a correct and stable results in Minkowski space. In the Section V we mention  some outcomes  for  continuation of the lattice gluon propagator and we conclude and discuss some open questions in Section VI.

\section{Analytical continuation I, PT-BFM gluon SDE}

In quantized gauge theory one needs to fix gauge in order to be able to calculate GFs. Usual approach is Fadeev-Popov \cite{FP} gauge fixing procedure which makes perturbation calculation particularly feasible, however   GFs become  gauge fixing dependent.   Pinch Technique is the S-matrix based construction of gauge fixing independent GFs. After reorganization of GFs (calculated at any gauge) one get new GFs which satisfy Ward Identities instead of usual more complicated Slavnov-Taylor identities. It has been proved to  all order of the perturbation that Pinch Technique is equivalent to the theory quantized by Background Field Method  at Feynman gauge. Guiding by the principles of Pinch Technique, the  PT-BFM gluon SDE has been recently studied in \cite{JHEP,cornwall2009}.

To get infrared finite gluon propagator (and running coupling), the gluon propagator must  loose its perturbative $1/q^2$ pole through the  Schwinger mechanism in Yang-Mills theory \cite{AGBIPA2011}. It underlies on the assumption of infrared singular gluon vertex  in a way it leaves polarization tensor transverse (gauge invariant). In this paper we simply use the derived PT-BFM equation in \cite{JHEP}, where Schwinger mechanism is employed through the simple Ansatz for the improper (two leg dressed) three gluon PT-BFM dressed vertex 
\be 
d(k)\tilde{\Gamma}_{\nu\alpha\beta}(k,q)d(k+q)=\int d\om \rho(\om)\frac{1}{k^2-\om+\ep}\Gamma_{\nu\alpha\beta}^L
\frac{1}{(k+q)^2-\om+\ep} +d(k)\tilde{\Gamma}_{\nu\alpha\beta}^Td(k+q)
 \, ,
\ee
where $\Gamma_{\nu\alpha\beta}^L$ satisfies tree level WTI and $d$ is scalar function related to the all order PT-BFM   gluon propagator  which in Landau gauge reads
\be
G^{\mu\nu}=\left[-g^{\mu\nu}+\frac{k^{\mu}k^{\nu}}{k^2}\right] d(k^2)
\ee
and satisfies generalized Lehman representation 
\be \label{spectral}
d(k^2)=\int d\om \frac{\rho(\om)}{k^2-\om+\ep}\, 
\ee
and $\tilde{\Gamma}_T$ is the rest of the three gluon improper vertex which is not specified by gauge invariance.
The essential feature of the vertex $\Gamma_{\nu\alpha\beta}$ is that apart the structure dictated by WTI  it also includes $1/q^2$ pole term which gives rise to infrared finite solution. For this purpose the following form 
\bea \label{YMSCHWINGER}
d(k)\tilde{\Gamma}_T^{\nu\alpha\beta}(k,q)d(k+q)&=&\int d\om \rho(\om)\frac{1}{k^2-\om+\ep}\Gamma_{\nu\alpha\beta}^T
\frac{1}{(k+q)^2-\om+\ep} \,
\nn \\
\Gamma^{\nu\alpha\beta}_T(k,q)&=&c_1[(2k+q)_{\nu}+\frac{q_{\nu}}{q^2}(-2k.q-q^2)]g_{\alpha\beta}
+[c_3+\frac{c_2}{2q^2}((k+q)^2+k^2))](q_{\beta}g_{\nu\alpha}-q_{\alpha}g_{\nu\beta}) \, ,
\eea
has been proposed in \cite{JHEP}. This vertex is transverse in respect to $q$ ($q.\Gamma=0$) and it  respects Bose symmetry to two quantum legs interchange as  its origin is due to the quantum loops. 
In this respect there are no singularities associated with external gluons, avoiding thus singular footprint in (however perturbative) any S-matrix element.

Finally, after the renormalization, it leads to the following form of linearized SDE in Euclidean space:
\bea \label{haf}
d_E^{-1}(q_E^2)=q_E^2\left\{K+bg^2\int_0^{q_E^2/4} dz \sqrt{1-\frac{4z}{q_E^2}} d_E(z)\right\}
\nn \\
+\gamma bg^2 \int_0^{q_E^2/4} dz z \sqrt{1-\frac{4z}{q_E^2}} d_E (z)+d_E^{-1}(0) \, 
\eea
where the second line arises due to the Ansatz for the gluon vertex (\ref{YMSCHWINGER}) and $K$ is the renormalization constant.
Thus the strength of the dynamical mass generation is triggered through  the  adopted coupling constants $c_1,c_2,c_3$  which is  fully equivalent to the  introduction of (in principle arbitrary)  constant $\gamma$ and infrared value $d_E^{-1}(0) $ (for completeness recall that in the paper \cite{JHEP} $d(0)$ has be calculated since one of $c_i$ was fixed by hand). 

The analytical assumption (\ref{spectral}) is the right key for  finding a correct continuation in Minkowski space. All the singularity structure is encoded in the distribution $\rho$. Thus if one  knows somehow the essential type of  singularity (the example is simple, i.e. $\rho=\delta(x-a)$ in this case) then one can estimate the  positions of induced  singularity structure, e.g. the position of the first branch points (thresholds are  examples generated by simple poles of internal propagators). Consequently , after making formal analytical continuation of SDE one can evaluate the imaginary and the real part of propagator function at domain where the function $d$ is continuous.  This procedure actually works numerically  for the models, where  the continuous function part of $\rho$ is smooth enough and associated principal value integrals can be performed with high numerical accuracy, see \cite{SAULI} for a review.  However when the coupling constant  is large  enough, one is faced to an oscillatory behavior of the spectral function $\rho$ \cite{SAUBI,SAULIJHEP,KUGO} and the SDE as an integral equation for $\rho$ becomes unstable. We found the system of such equations extremely unstable in the case of PT-BFM gluon SDE as well and due to this reason we develop different strategy for this purpose.  

%\begin{figure} 
%\centerline{\epsfig{figure=curve.eps,width=8truecm,height=6truecm,angle=0}}
%\caption[r]{\label{contour} Integration contour in complex plane of momenta.}
%\end{figure}

Due to a simple linear structure of the SDE (\ref{haf}) the analytical continuation is very straightforward. Recall  the definition here 
\be
d_E(q_E^2)=-d(-q^2) \,\, {\mbox{for}} \,\,  q^2<0 \,\, ,
\ee
where now $d$ is Minkowski space PT-BFM gluon propagator.  Putting $q^2_E \rightarrow -q^2$ and considering positive (timelike) $q^2$ continuation we  get the SDE  for continued gluon propagator:
\bea \label{haf2}
d^{-1}(q^2)=q^2\left\{K+bg^2\int_0^{q^2/4} dz \sqrt{1-\frac{4z}{q^2}} d(z)\right\}
\nn \\
+\gamma bg^2 \int_0^{q^2/4} dz z \sqrt{1-\frac{4z}{q^2}} d (z)+d^{-1}(0)+i\epsilon \, \,
\eea
where  the substitution $z\rightarrow -z$ was made and  also the both sides were multiplied by $-1$.

While the Eq. (\ref{haf2}) is formally identical to its spacelike counterpartner , the main difference should be stressed:  The function $d$ on the both sides is complex function, the one on the  rhs. of SDE is integrated over the region where all  singularities are located. Meaning of Feynman  $i\epsilon$ is as usual, it is responsible for the  generation of absorptive part of $d$ at the timelike region, however here,   we keep it small and finite in our numerical treatment.    
Having $d$ complex, the equation (\ref{haf2})  represents two coupled real equations for an imaginary and the real  parts of $d$. Thus  the presence of small imaginary is crucial for obtaining a correct Minkowski solution. Clearly switching off
imaginary factor we would repeat   Euclidean space solution again, albeit now for timelike $q^2$. 
Clearly such solution do not complete the full solution in Minkowski space since the timelike solution of Eq. (\ref{haf2}) is purely real and  contradicts the assumption made. Also note, such solution has an infrared  discontinuity 
$d^{-1}(0^+)-d^{-1}(0^-)=2\Lambda^2$ (left limit is the Euclidean) , while for a continuous solution we need to adjust $d^{-1}(0^+)=-\Lambda^2$. At this place, we can mention we have found a real solutions of Eq. (\ref{haf2}) for $d^{-1}(0^+)<0$ , which includes the case
 $d^{-1}(0^+)=-\Lambda^2$ as well. Up to the later case all  of them posses second order  discontinuity, while for the choice $d^{-1}(0^+)=-\Lambda^2$ we also got the real solution but with the first order singularity only.
 Of course, these real solutions cannot be interpreted as the analytical continuation which we are looking for since the reality of $d$  completely contradicts our assumptions.   
 Much interesting for us is the following observation: for mentioned solutions we  get    the second and the first order singularity respectively, which 
is  located at $q^2=0$. It is the motivation to look, if not yet an evidence, for the complex solution with the branch point located at the origin of $q^2$ complex plane.   

We argue here that the solution of Eq. (\ref{haf2}) is not unique, there must be more solutions in addition, at least one is complex and  respects Lehman representation and corresponds thus to the  continuation to the Minkowski timelike domain of $q^2$, this later has  the branch point located at the origin of Minkowski $q^2=0$. There exist probably more (infinite number is not excluded) solutions in  other Riemann sheets if $d$ is a multivaluable analytical functions. The required solution is represented by the function which starts to be complex  from the origin in accordance with the introduction of trivial  boundary in the  integral expression (\ref{spectral}). Actually we got numerical  evidence for such  statement  by our numerical search. 

Before discussing more  details,  we improve the linearized SDE in a way it gives correct UV asymptotic behavior, for which case we also present the numerical solution. 

\section{Renormalization group improved  gluon SDE}

 Solving the SDE perturbatively by taking $d(z)\rightarrow 1/z$ one  get the correct 1 loop perturbative limit
\be
d(x)|_{pert}\rightarrow \frac{1}{x(1+bg^2 ln(x/\mu^2)} \, .
\ee
However, when solving SDE selfconsistently   then the solution of  (\ref{haf2}) leads to the slower decrement with growing momenta $q^2$, instead of log suppression we would get $(1+bg^2 ln(x/\mu^2))^{1/2}$ (see \cite{JHEP} for the details) behavior of the gluonic form factor, which  is believed is  the  prize of the gauge technique and associated renormalization simplifications.

In order to restore the correct RG behavior of the SDE solution one needs to replace 
\be  \label{subst}
g^2 d(z)\rightarrow\frac{g^2}{g^2_{UV}(z)}d(z)  
\ee
in the integrands of SDE, where $g^2_{UV}(z)$ is some function with  the perturbation theory running coupling ultraviolet asymptotic.

To accomplish  this we take the function $g_{UV} $ in the form
\be
g^2_{UV}(z)=\frac{1}{b ln(e+\frac{z}{\Lambda^2})}.
\ee  
 Up to an required asymptotic it is smooth function enough, which ensures   unwanted  infrared modification of the SDE kernel.

As a first necessary step  we obtain  the solution of PT-BFM SDE in the spacelike region and we solve the SDE in the Euclidean space. Incorporating substitution (\ref{subst}), fitting the renormalization constant $K=1$ we solved the following equation 
\bea \label{haf3}
d_E^{-1}(q_E^2)=q_E^2\left\{K+b\int_0^{q_E^2/4} dz \sqrt{1-\frac{4z}{q_E^2}} d_E(z)\frac{g^2}{g^2_{UV}(z)}\right\}
\nn \\
+\gamma b\int_0^{q_E^2/4} dz z \sqrt{1-\frac{4z}{q_E^2}} d_E(z) \frac{g^2}{g^2_{UV}(z)}+d_E^{-1}(0) \, .
\eea

\begin{figure} 
\centerline{\epsfig{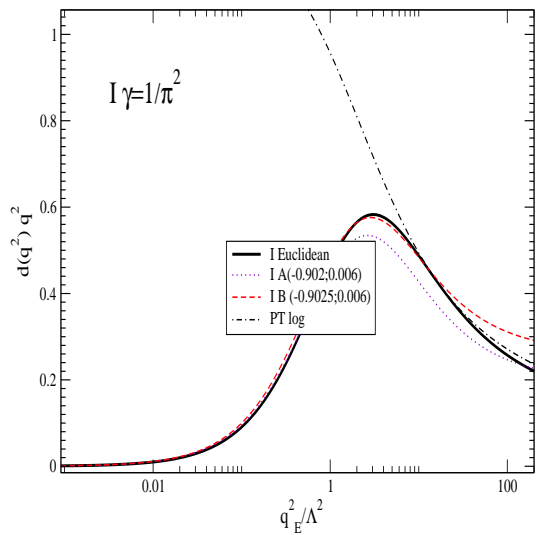}}
\caption[caption]{\label{propagII} Spacelike solution for the model with coupling $\gamma=1/\pi^2$ and $d^{-1}(0^-)=\Lambda^2$. Comparison with spectral representation fits are added,  the bracket shows the numerical fit:   
$( Re d(0^+)^{-1},\epsilon)$ (in units of $\Lambda^2$) for which the Minkowski SDE has been solved.\vspace{2cm}}
\end{figure}

As in the previous case of Eq. (\ref{haf2}), two parameters stay completely free in given truncation. For simplicity and quite naturally we take the infrared gluon mass  identical with $\Lambda_QCD$. 
\be
d^{-1}(0)=\Lambda^2 \, .
\ee
and then looking what happens when $\gamma$ is varied. We have found there is  little dependence of the solution even when we change  $\gamma$ about order of magnitude, for the solutions see Fig. \ref{propagII}.

\begin{figure}
\centerline{\epsfig{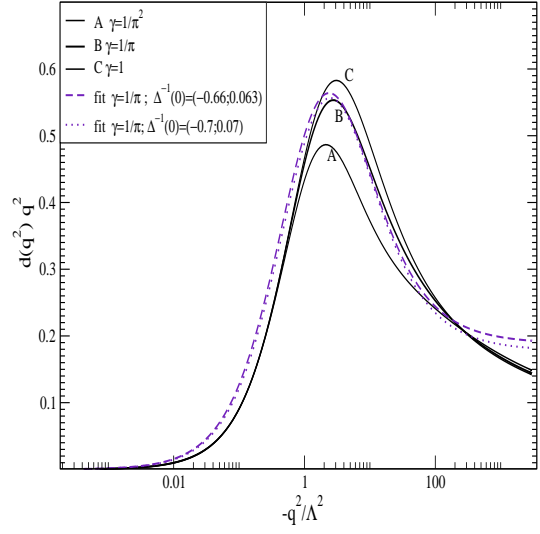}}
\caption[caption]{\label{secondfit} Spacelike solution for various  $\gamma$.  Dashed and dotted lines represent  two fits obtained for   $-\gamma=1/\pi$ from Minkowski SDE solution (see the text for the details). The fits are labeled by  values of  $ Re d(0^+)^{-1},\epsilon$ in units of $\Lambda^2$.\vspace{2cm}}
\end{figure}

After a slight rearrangement, the RG improved PT-BFM SDE (\ref{haf3}) has been solved by the iteration.  The iteration procedure  shows up very fast convergence, providing 20-30 iteration steps are enough to achieve vanishing difference between the  two followed iterations. Since  the Euclidean  correlator    is quite  smooth function thus the resulting solutions are numerically stable.  Thanks to the 1-dimensionality of the problem we can use a grid with large number of mesh points giving easily 3 to 6 digit accuracy  without a special treatment of the upper boundaries for which we use simple  step function.  In our treatment we do not define a new gluon form factor neither gluon dynamical mass as this cannot be done without ambiguity.
 In figure \ref{propagII} we plot the resulting solution  for the dressing function $F=q^2d$. The  comparison  with the 1-loop perturbative running coupling asymptotic is shown in figure \ref{secondfit}.

\section{Analytical Continuation II, Results for Renormgroup Improved PT-BFM SDE}

If gluon were a stable massive particle we could get delta distribution $\delta(\omega-\mu^2)$ as a  first substantial singularity in gluon spectral function. Further there could  be a  second singularity  at $4\mu^2$ where the continuous part of spectra usually starts. As the massless ghosts were omitted we could get further singularities at $9\mu^2,16\mu^2,...$  all these singularities exhibit ourself as a finite cusps in the real part of $d$, since $\rho$ has no derivatives at these points and all of them  could be located  on the cut in complex plane which is identical with the real positive semiaxis of Minkowski $q^2$.

However, instead of dealing with stable vector particle, we are dealing with confined gluon field.   
Let us imagine a process when "high energy" gluon is emitted with some virtuality $q^2$. It starts to loose its energy  through the pair creation and bremsstrahlung of soft gluons, all these colored intermediate states   are finally (and  simultaneously) neutralized in a color singlet glueballs (hadrons, when real QCD  with quarks is considered). Such color neutralization can happen  only through the  interaction with
other colored partons- gluons and quarks. Gluon never escapes as a free on shell particle, no matter what the typical gluon mass scale would be. From this picture  it is apparent: the timelike structure can be quite complicated as many "intermediate states" contributes significantly because of  strong coupling.  The emission and absorption of quanta of the fields are related with threshold singularities when unconfined particles are considered. Here, for the case of gluons, we expect something similar, but perhaps qualitatively different: the absorption and emission of confined object could reflect their short life $t\simeq 1/\Lambda_{QCD}$ ,  we expect this reflection is encoded in the  singularity structure typical for confinement.  Thus considering spectral function  one can really  expect kind of cusps instead of delta function and some more smooth non-monotonic behavior instead of threshold cusps. Constrained by construction here, we do not expect their positions in a complex plane away from a positive real axis. 

In any case, assuming spectral function has a continuous part we must get standard dispersion relation for
\bea \label{hurvajz}
Re d(k^2)&=& P. \int d\om \frac{\rho(\om)}{k^2-\om}\, ,
\nn \\
Im d(k^2)&=& -\pi \rho(k^2)\, ,
\eea 
where P. stand for principal value integration.

Without knowledge of branch points and singularities  in $\rho$ one is faced with problem of  performing  limiting analytical continuation to the borderline of analyticity- to the cut, where at (hopefully) isolated points  we assume $d$  has aforementioned singularities of not yet specified type. Not necessarily but likely, they  can be related with a branch points of multivaluable analytical function and very likely the number of them increases with growing $q^2$. It is our assumption that all the information about singularities is correctly captured in Minkowski space continued SDE for gluon. Of course , one can expect that various approximation, which are necessary complements of given  truncation of SDEs system has an impact on the solution, e.g. on the position, number of branch points and even the analyticity can be destroyed. Finally remind the well known examples:  this is the Dirac delta function in spectrum which implies pole in the propagator, similarly Heaviside step function   $\theta(x-a)$ gives $log (q^2-a)$ divergence and for instance $\theta(1-T/x)\sqrt(1-T/x)$ gives usual particle production threshold at $T$. That it is  easy to write down a plethora of arbitrarily  moderated singularities, which are not familiarly known from  particle physics is quite obvious.

In previous discussion we offer some arguments which are crucial for the method of the solution. 
Let us summarize and specify our assumptions to the following points:
1) Solution for a complex arguments $q^2, Im \, q^2 \not = 0$ is given by analytical continuation of Euclidean solution.
2)  Minkowski solution corresponds with $Im \, q^2=0$, which is in principle obtainable by limiting procedure based on (1), can be directly obtained from  continued  Euclidean SDE, i.e. singularities are integrable and isolated.
3) The first singularity of the gluon propagator is located at the origin of complex plane.

Continuation of PT-BFM gluon SDE is straightforward as there is only function  $g$ added when compared to the previous case. First let us rewrite Euclidean SDE by using  Minkowski conventions $q^2_E=-q^2$

 Following  the convention $q^2_E=-q^2$ for negative spacelike Minkowski $q^2=q^2_0-{\bf{q^2}}$
then we can rewrite the Eq. (\ref{haf3}) like
\bea \label{haf4}
d_E^{-1}(-q^2)=-q^2\left\{K+b\int_0^{-q^2/4} dz \sqrt{1+\frac{4z}{q^2}} \frac{d_E(z)}{g^2_{E}(z)}\right\}
\nn \\
+\gamma b\int_0^{-q^2/4} dz z \sqrt{1+\frac{4z}{q^2}}\frac{d(z)}{g^2_{E}(z)}+d_E^{-1}(0) \, ,
\eea
which is valid  for negative $q^2=q^2_0-{\bf{q^2}}$.
The continuation $d(q^2)=-d_E(-q^2_E)$ to positive $q^2$ then reads
\bea \label{haf5}
d^{-1}(q^2)=q^2\left\{K+b\int_0^{q^2/4} dx \sqrt{1-\frac{4x}{q^2}} \frac{d(x)}{g^2_{c}(x)}\right\}
\nn \\
+\gamma b\int_0^{q^2/4} dx x \sqrt{1-\frac{4x}{q^2}}\frac{d(x)}{g^2_{c}(x)}+d^{-1}(0)+i\epsilon \, ,
\eea
i.e. now the integral domain on the rhs. of (\ref{haf}) becomes purely timelike as the all arguments appearing in SDE  are.
 Again we have  performed the substitution $x=-z$ in order to show formal similarity with the Euclidean equation. The function  $ g_c$ in Eq. (\ref{haf4}) should be  the analytical continuation of our UV renormgroup improver, for which we take the function
\be \label{snyoropuse}
g^2(z)=1/b ln(e+|q^2|/\Lambda^2)
\ee
where $e=2.73..$ and  we neglected $i\pi$ term as a legitimate simplification valid at ultraviolet. 

Here we briefly mention trivial fact that the Eq. (\ref{haf5}) includes unphysical solutions, e.g. the real solution 
$ d(z)= d_E(z) $, which corresponds to the choice $d^{-1}(0)=d_E^{-1}(0) $.
We have also found some real solutions even for more "correct" choice    $d^{-1}(0)<0$.
Again, they are clearly not hot candidates for wanted analytical continuation, as they do not agree with the dispersion relation (\ref{spectral}).

We solve the  Eq. (\ref{haf4}) numerically at positive semiaxis $q^2=(0,+\infty)$ by the method of iterations.
We were searching for complex $d$ which can have nonzero imaginary part everywhere at $R^+$.
Numerically we take $\epsilon$ as constant satisfying  $\epsilon<<|d(0^+)|$ for all values of $q^2$.
Numerically, $\epsilon$ must be taken nonzero in order to accurately perform numerical integration in the vicinity of
branch points in $d$.  A bit unexpectedly a wide variety of stable numerical solutions of Eq (\ref{haf4}) has been found.

We anticipate here, the main feature of the solution is the appearance of two relatively large peaks with mutually opposite signs (see the results in figures 3-5) .  Such singularities and the associated oscillations make the appropriate  numerical findings a really hard problem. We achieve stable numerics by using Gaussian integrators such that intercept between two neighborhood points is smaller then $\epsilon$ in the vicinity of singularities. We found advantageous to iterate not the function $d$, but rather its inverse which leads to very fast numerical convergence. Quite independently on the initial guess, typically $\simeq$ 30-40 iterations are enough to get vanishing $ (10^{-15})$ difference between last iterations. For instance the curves  shown in Fig 2. were evaluated for  9000 Gaussian mesh points.

In Euclidean space, the value of BFM gluon propagator at zero momenta represents a free parameter which must be fitted by hand. The value $d_E(0)=\Lambda$ was chosen for simplicity. There is no free choice when dealing with the SDE in Minkowski space and the right limit of $d$, ie. $d(0^+)$  value is necessarily  constrained by global analytical property of $d$ (stress here , we assume $q^2=0$ is a branch point, in principle we allows the left and the right limit differ). Here we describe the numerical treatment  which has been actually used in this paper.  
  
To find the best analytical continuation, we scan the values $d(0^+)$ and $\epsilon$, such that  $\epsilon/d(0^+)<<1$ and
$d(0^+)\simeq (0^-)=\Lambda$, for each set we  find the solution of Minkowski SDEs. 
 Then we check the dispersion relation for the solution $Re d(q^2)$ and $Im \, d(q^2)$. 
The one which best  reproduce the Euclidean solution via assumed dispersion relation for $d$ is called the best analytical continuation (in sense of the point (2)). 
In accordance with spectral representation one expects
vanishing imaginary part at the infrared as the correct analytical continuum limit. 
We can conclude, that within an accuracy limited by finite size of $\epsilon$  one can always find the solution of timelike SDE.  We expect the numerical accuracy of SDE solution is driven by the ratio, i.e  $\epsilon/\Lambda\simeq 0.1-1\%$, while 
the  agreement with assumed analyticity is a different issue. 
For this purpose we qualify the quality of analytical continuation by evaluating  the following difference:
\be \label{dif}
\sigma(q^2)=q^2_E\left(d_E(q_E^2)-\frac{1}{\pi}\int d\om \frac{-Im d(l^2)}{q^2_E+l^2}\right)\, 
\ee
where $d_E$ is the numerical solution of (\ref{haf3}) and $d$ is the numerical solution of (\ref{haf5}).
The best analytical approximation was achieved by minimizing $\sigma$ and thus getting a few percentage deviation (for almost all $q^2$) at the minimum.  For instance, for the case of small Schwinger coupling $\gamma=0.1 $,  the best analytical infrared fit we were able to trigger numerically  uses $\epsilon=10^{-3}\Lambda^2$ while keeping  $Re \, d(0^+) \simeq d_E(0)=\Lambda^2$. Let us stress the estimate of the numerical systematical error $\simeq  10^{-3}$ (for absolute values of $d$) is always much smaller that the deviations from analyticity which can be estimated  approximately by  
$(d_E(0)-d(0^+))/(d_E(0)+d(0^+)) \simeq 0.1 $ in the infrared.

\section{$\gamma$ dependence of Minkowski solution}

The parameters $\gamma$ represents the strength of the effective coupling constant which arise due to the Schwinger mechanism.  Thus numerical value of $\gamma$ should be  responsible for the size of $d(0)$ which  can be estimated by the feedback of SDE solutions. Therefore we consider several 
values of $\gamma$ and report numerical results here for small $\gamma=1/\pi^2$ and "large" $\gamma=\pm 1/\pi$ couplings respectively.

In previous section we described the search of continued solution in details. In  all studied cases we have found that the agreement with the exact analyticity is somehow limited ( $\sigma=0$ in the exact case).
For a larger  $\gamma $  it was difficult to get stable solution and minimize $\sigma$ so we were limited by  concerning  values $|\gamma|<1$.   
 A few examples of a backward analytical continuation are shown in figure \ref{propagII} for $\gamma=1/\pi^2$ and in figure \ref{secondfit} for $\gamma=-1/\pi$,    where they are compared with the original Euclidean solutions. Following the numerical procedure described, we label various solutions  by  the pair of numbers $(Re\,  d^{-1}(0^+),\epsilon)$ in the units of $\Lambda$. The difference between two initial choices $(-0.902;0.006)$ and $(-0.9025;0.006)$ exhibits numerical sensitivity in the case of small $\gamma$.

While the dependence on $\gamma$ is rather mild in the case of the Euclidean solutions,  we have found that this is not the case of the Minkowski space solutions. The solutions show up several  cusps, number of them and their shapes largely change when varying $\gamma$. The only approximate invariant is the fact that the first maximum is positive
(in sense of spectral function $\rho$) and located more or less at $\Lambda$, while the second peak is negative
with the position  shifted slightly to a higher scale. Other local maxima and minima are generated as well, also the discontinuity in the first derivatives are found. The first peak is located at scale  $\Lambda$, which is in fact our  choice of the infrared value of the gluon propagator, and quite expectingly, when the it is sharp enough,   then one can see associated quasithreshold (at the scale $2\Lambda$). The second peak is a surprise of the Minkowski solution. For a large momenta $q^2>>\Lambda^2$ the second singularity regulates the contribution from the first one. Such screening 
is actually very large for small Schwinger coupling Considering  the approximation $\rho(\omega)=\delta(\omega-\mu_1^2)-\delta(\omega-\mu^2)$ of the infrared spectral function, the screening of the first peak by the second one is obvious. For instance for  small $\gamma=1/\pi^2$ we have found the position of the peaks are $\mu_1\simeq \Lambda, \mu_2\simeq \sqrt{2.5}\Lambda$.

In order to get some feeling  how the numerical solution depends on the details which are not  theoretically fully determined, we  modify two parameters in the solely Minkowski SDE (\ref{haf5}), then after getting the solution we again use dispersion relation for $d$ and see how the Euclidean solution is recovered.
The main purpose to do this is to see whether the approximation employed has crucial influence on the analyticity.   As a first, we have changed the sign of $\gamma$ in  the Minkowski SDE and then compare obtained solution with the original Euclidean one via backward continuation (\ref{dif}).
However alternating sign of $\gamma$ is ad hoc here, there exists a good motivation: beyond our truncation the constant $\gamma$ will be replaced by the product of vertex function for   glueball-like BSE for the gluon vertex. We expect , such product substantial changes  when going to Minkowski timelike axis- it becomes complex and very likely enhanced there- therefore alternating the sign may not be so radical for our rough estimate.  Actually, as we have found the deviations from analyticity $\sigma$ reaches smaller minima,  such scenario is possible and perhaps preferred when compared to the case when $\gamma$ is unchanged.    
The resulting numerical solution are labeled as Model II in all presented figures, while what is labeled by letter I is calculated without any changes. Also the fits in figure 2 are based on the backward continuation of Minkowski solutions II which are displayed in figures 3-5.  

Secondly, we slightly change  the  improver $g_c$, again solely in Minkowski SDE.  For this purpose we  replace the Euler constant $e$  in the expression (\ref{snyoropuse}) by free parameter $C$. As a result, we have found $g_c $ enhanced (taking $C=2.5$) is numerically preferred for positive value of $\gamma$, while taking $g_c$ slightly suppressed (taking $C=3.5$) we can get better fits of Euclidean solution for negative $\gamma$. As an effect these changes modify the shape of the peaks.  The appropriate solution for $\gamma=1/\pi$ coupling is  added in   figure \ref{vysledek}. As one can see the second peak quite changes not only position but  it transforms to a much narrower cusp. From this one can conclude that the analytical form of the resulting Minkowski propagator can be quite sensitive to the  numerical approximations, which  can be only marginal in the Euclidean treatment. 

As one can see from the Figs. 4 and 5, taking $\gamma $ small one get the solution with two very narrow peaks located at few $\Lambda^2$.  We could expect these peaks should be visible  in a hadronic S-matrix element whenever timelike gluon is interchanged as an intermediate excitation. There would be "long living gluon" and "long living gluonic ghost" at the scale $\Lambda$ according to their numerical "widths" which has the size by order smaller then $\Lambda$. Disregarding the reality of such scenario, we get interesting solution-two possible  excitations appear for a single  field $A$. Let us imagine what happens when we collide hadrons in such a world.  Whenever energetically possible  the  quarks (and gluons itself)
would preferably exchange low energy gluons with positive virtuality corresponding to the singularities or  maxima positions  and such peaks of  gluonic colored corellator will become visible through the  color subchannels. Even  after the integration over the distribution of quarks, such gluonic colored mediators would be  observed as resonances with very large cross sections (although without any  doubt we know that many gluon interchanges are necessary to describe hadronic amplitudes, as only very large number can be responsible for formation of hadronic resonances, for very nice illustration see the study \cite{MARCOT2002} on $\pi-\pi$ scattering). 

Increasing $\gamma$ changes the shape of the gluon propagator, especially  the sharp peaks are melted down as they are transformed into more smooth  maxima. For large $\gamma$ this is the branch point in the origin which is sole numerically observed singularity of the gluon propagator.   The details of melting peaks is nicely seen in  figure \ref{check4}.

\section{Fitting the lattice}

The framework of PT-BFM SDEs is based on  well defined theoretical background and in principle it could finally provide a gauge fixing independent set of new Greens functions. Nevertheless, the quantitative comparison with the standard-gauge fixed lattice simulation is further stimulating. Actually, there is small quantitative difference between decoupling Euclidean solutions, and the lattice evaluation of gluon propagator in Landau gauge \cite{AGBIPA2008,AGBIPA2010,BI2010}. In this respect, the analytical continuation obtained from PT-BFM SDE can be regarded as an reliable estimate of Minkowski space continuation of lattice data. 

Based on suitable numerical fits at Euclidean regime, a various form of continuation were proposed for scaling solutions of SDEs \cite{Alkofer}. According to the conclusion and numerical findings in \cite{Alkofer}, the  singularities of scaling solution studied are located near the real timelike semiaxis of the momenta. It is quite likely that the qualitative analytical properties of various solutions of YM SDEs systems are frankly insensitive to finiteness of $d(0)$ and thus  both, the scaling and the decoupling solutions of gluon propagators can be captured by the formula for Lehman representation.
Then the quantitative difference will result from different forms of $\rho$, whatever it could be. Recent lattice data show finite infrared gluon propagator, which however is usual property of massive propagator, but as we have shown it  does not necessary mean propagation of massive particle. There are legitimate attempts to move the first branch point
 towards the higher timelike scale \cite{SUGANUMA,IRITANI}, which fits are  based on low $q^2$ lattice data. The first singularity fitted in this way assume quite obscured form of the spectral function. It is taken as a product of delta function and the inverse  square root function singular at the point.  Although it allows a reasonable estimate of the singularity position $m=600MeV$ and it provides rough numerical  fit of the Euclidean data at the region 0.5-5 GeV , the choice of the singularity form is far from obvious. The other lattice fits were proposed in the literature. For instance 
in the papers \cite{DUGRSO2008,DUOLVA2010} the function
\be \label{dudalfit}
\Delta(k^2)=\frac{k^2+M^2}{k^4+(M^2+m^2)k^2+\lambda^4} \,
\ee
was used to describe the lattice prediction for the infrared gluon propagator up to $k \simeq 1.5 GeV$.
The function  $\Delta$ has two complex conjugated poles and thus certainly does not belong to the sort of functions satisfying dispersion relation (\ref{spectral}). However, as it was shown recently in \cite{DUDAL2011}, when $\Delta$ is considered in the loop, the resulting expression can be cast into the form of Eq. (\ref{spectral}) (in fact, it is  kind of an evidence, that albeit $D$ does not have Lehman representation, its inverse has a part, which after the renormalization can be expressed through the  dispersion  relation).

In our case, we have an additional hint  estimated from the behavior  of SDEs solution: the PT-BFM   gluon propagator  posses the first branch point at $q^2=0$ and the others are generated on the real axis by the Gauge Technique construction and  selfconsistently determined by solving the continued SDE (\ref{haf5}). Making backward analytical continuation we reproduce the Euclidean solution approximately.  From this we know, that we did not cross any singularities when one deforms usual Wick rotation contour  at $q_0$ plane. This is a strong argument
which supports our findings, when comparing for instance to ad hoc guess of the singularity structure  from fitting of  shape of the Euclidean lattice data. 

\begin{figure} 
\centerline{\epsfig{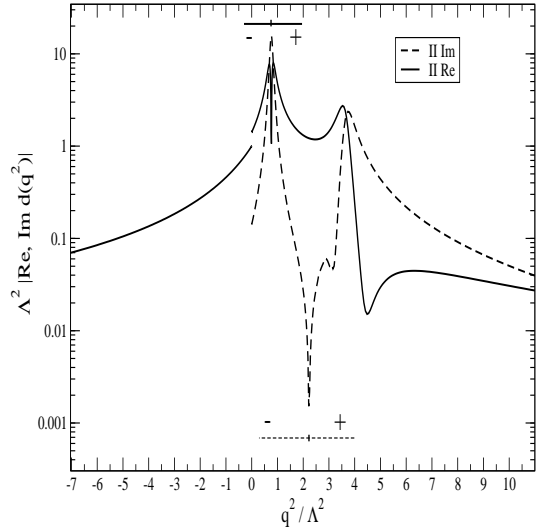}}
\caption[r]{\label{vyslin2} Magnitude of PT-BFM gluon propagator in whole Minkowski space for $\gamma=1/\pi$ and  $Re d(0^+)^{-1}=0.7\Lambda^2$, $Im d(0^+)^{-1}=0.07\Lambda^2$, $d^{-1}(0^-)=\Lambda^2$).\vspace{2cm}}
\end{figure}

\begin{figure} 
\centerline{\epsfig{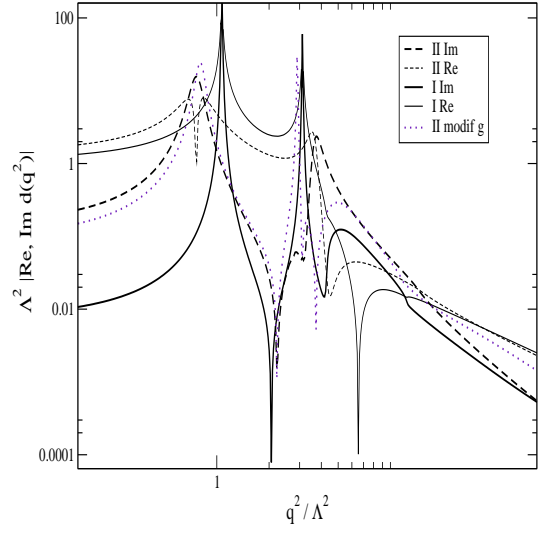}}
\caption[caption]{\label{vysledek} Magnitude of PT-BFM gluon propagator in  Minkowski space for Model I: $\gamma=1/\pi^2$ and  $Re d(0^+)^{-1}=0.9\Lambda$, $\epsilon=0.006\Lambda^2$,  corresponding $d^{-1}_E(0)=\Lambda^2$ and the Model II, which is the same as in the Fig. \ref{vyslin2}. Dotted lines represents Model II but with the function $g_c$ as described in the text.\vspace{2cm} }
\end{figure}

\begin{figure} \label{petka}
\centerline{\epsfig{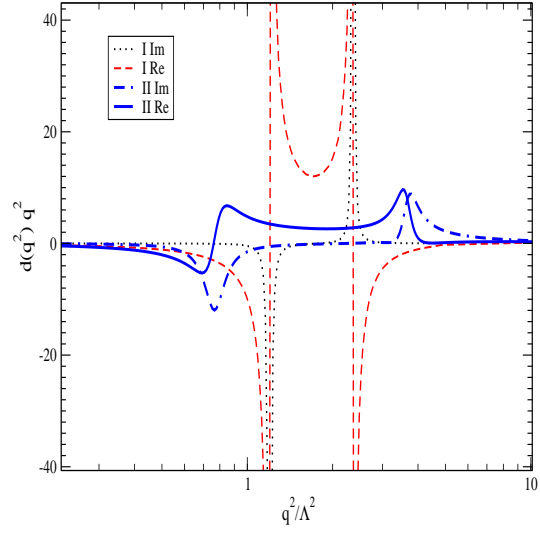}}
\caption[caption]{\label{check4} Details of the singularities (and shape) of the timelike gluon form factor as obtained for small $\gamma=1/\pi^2$ (Model I) and moderate coupling $\gamma=1/\pi$ (Model II) respectively.}
\end{figure}

\section{Conclusion}

The main purpose of our paper is to show that Stieltjes transformation or equivalently generalized Lehmann representation exists for PT-BFM gluon propagator. Very likely, the same applies  for the conventional Landau gauge gluon propagator recently seen in the lattice simulation. We have found reliable spectral function for all $q^2$, not only at the infrared or ultraviolet part separately. According to the expectations, the gluon propagator exhibits violation of positivity. In accordance with confinement the main singularity does not correspond with single delta function. However, for a small coupling characterizing the strength of the YM Schwinger-mechanism the infrared gluon propagator can be approximated by two mutually opposite sign  Yukawa  propagators, while the whole propagator function needs a certain amount of the continuous background in addition.
Increasing $\gamma$ we get a more realistic picture of confinement -the sharp peaks in spectral function are melted down, while 
the Lehman approximation is less approximative, perhaps showing the evidence for  complex conjugated poles of the from (\ref{dudalfit}).
 According to our approximative findings, the continuous spectrum starts at zero momenta, where we expect the first branch point.
For small $\gamma$ , this  is in excellent agreement with the assumption: the propagator is finite and purely real at light cone $q^2=0$.
However for a larger $\gamma$ we get small imaginary part, which contradicts coexistence of Lehman representation and finiteness of the real part of gluon propagator at zero momenta. This, although small discrepancy tell us that our analytical assumption is not  fulfilled , but is justified as an approximation only.

In order to discuss a possible sources of this discrepance let us tentatively divide the gluon PT-BFM propagator into two peaces:
\be  
d(q^2)=\int d\om \frac{\rho(x)}{q^2-x+\ep}+r_{NA}(q^2) \, ,
\ee
where $r_{NA}(q^2)$ should be small when compared to the full $d(q^2)$ and we distinguish among several possibilities.

1.  where $r_{NA}$ is the marginal complex remnant which does not 
obey Stieltjes representation  but does not contradict Wick rotation and thus the primacy of Euclidean calculation as well.
The example is complex pole located at point $z$ with $Re x>0$ and  $Im z<0$ 

2. $r_{NA}$ does not allow Wick rotation, there are singularities and cuts in complex plane of $q^2$ (and $q_0$)
which makes the theories in Euclidean and Minkowski space different. In this case, the lattice data should be regarded as a certain approximation of Minkowski spacelike correlators. The example of such behaviour we  can mention is the function given by (\ref{dudalfit}), if it has  complex pole at some $z, Re z>0,Im z>0$.  

3.  $r_{NA}$ vanishes for the full solution. It is possible  it is a consequence of approximation employed, eg. it is a possible artifact of weakness of the Gauge Technique linearization, truncation of SDE system, modeling  Schwinger mechanism and modeling UV RG improver $g_c$. In this case the idea of analytical effective charge finds new applications beyond its original perturbative conjecture \cite{SHIRKOV,MISO1997,MISO1998,SHISOL2007} 

To distinguish among above three points  is beyond our recent analyzing power, however in future study one should always ask, what improvement of the knowledge has been achieved in this respect.

In practice, most of  theoretical description of QCD processes is based on the factorization schemes where the hard part of amplitudes is calculated by using perturbation theory while the non-perturbative information is included in the partonic distribution functions. 
Therefore the first hint we can get, can be obtained from the phenomenological studies which adopt the  running charges constrained by finite BFM gluon propagator, for  recent calculation of the proton structure function  see \cite{NATALE}, and \cite{NATALE2} for heavy meson decay study. The timelike gluons exchanged certainly appear in the s-channel 
of meson scattering or heavy meson strong decays. The calculation difficulty of such process is obvious,  the observed gluon peaks will interfere not only with itself, but with the first colorless resonances, e.g. $\rho$ meson which always arise nonperturbatively when very large number (infinite) gluons are exchanged    (\cite{MARCOT2002}). 
Nevertheless the difficulty of the problem, we believe that propagator calculated here can be useful when one would like to investigate glueball and hadron spectra in cases when timelike kinematics is especially pronounced.  For this purpose  inclusion of the quark fields remains to be done.

\end{document}